\documentclass[12pt]{article}%
\usepackage{amsmath,latexsym}
\usepackage{graphicx}
\usepackage{amsmath}
\usepackage{amsfonts}
\usepackage{amssymb}%
\begin{document}

\title{{A note on the cosmological constant in $f(R)$ gravity}}
   \author{
Peter K.F. Kuhfittig*\\
\footnote{kuhfitti@msoe.edu}
 \small Department of Mathematics, Milwaukee School of
Engineering,\\
\small Milwaukee, Wisconsin 53202-3109, USA}

\date{}
 \maketitle

\begin{abstract}\noindent
The starting point in this note is $f(R)$
modified gravity in a cosmological setting.
We assume a spatially flat Universe to
describe late-time cosmology and the
perfect-fluid equation of state
$p=\omega\rho$ to model the hypothesized
dark energy.  Given that on a cosmological
scale, $f(R)$ modified gravity must remain
close to Einstein gravity to be consistent
with observation, it was concluded that
either (1) Einstein's cosmological constant
is the only acceptable model for the
accelerated expansion or (2) that the
equation of state for dark energy is far
more complicated than the perfect-fluid
model and may even exclude a constant
$\omega$.\\

\noindent
Keywords\\
Cosmological Constant, $f(R)$ Gravity\\

\end{abstract}

\section{Introduction}

The discovery that our Universe is undergoing
an accelerated expansion \cite{aR98, sP99} has
led to a renewed interest in modified theories
of gravity.  One of the most important of these,
$f(R)$ modified gravity, replaces the Ricci
scalar $R$ in the Einstein-Hilbert action
\begin{equation*}
  S_{\text{EH}}=\int\sqrt{-g}\,R\,d^4x
\end{equation*}
by a nonlinear function $f(R)$:
\begin{equation*}
   S_{f(R)}=\int\sqrt{-g}\,f(R)\,d^4x.
\end{equation*}
(For a review, see Refs. \cite{SF10,
NO07, fL08}.)

An alternative to the modified gravity model
is the hypothesis that the acceleration is
due to a negative pressure \emph{dark energy},
implying that $\overset{..}{a}>0$ in the
Friedmann equation
\begin{equation*}
   \frac{\overset{..}{a}(t)}{a(t)}
   =-\frac{4\pi}{3}(\rho+3p).
\end{equation*}
(We are using units in which $c=G=1$.)  In
the equation of state $p=\omega\rho$,
$\overset{..}{a}>0$ corresponds to the
range of values $\omega <-1/3$, referred
to as \emph{quintessence dark energy}.
The case $\omega =-1$ is equivalent to
assuming Einstein's cosmological constant.
It has been forcefully argued by Bousso
\cite{rB12} that the cosmological constant
is the best model for dark energy.  In this
note we go a step further and propose that
$f(R)$ modified gravity implies that
$\omega =-1$ is the only allowed value in
the equation of state $p=\omega\rho$.
%END OF SECTION

\section{The solution}

For convenience of notation, we start with the
spherically symmetric line element
\begin{equation}\label{E:line1}
ds^{2}=-e^{2\Phi(r)}dt^{2}+\frac{dr^2}{1-b(r)/r}
+r^{2}(d\theta^{2}+\text{sin}^{2}\theta\,
d\phi^{2}).
\end{equation}
It was shown by Lobo \cite{LO09} that under the
assumption that $\Phi'(r)\equiv 0$, the Einstein
field equations are
\begin{equation}
   \rho(r)=F(r)\frac{b'(r)}{r^2},
\end{equation}
\begin{equation}
   p_r(r)=-F(r)\frac{b(r)}{r^3}+F'(r)
      \frac{rb'(r)-b(r)}{2r^2}
          -F''(r)\left(1-\frac{b(r)}{r}\right),
\end{equation}
and
\begin{equation}
   p_t(r)=-\frac{F'(r)}{r}\left(1-\frac{b(r)}{r}\right)
   +\frac{F(r)}{2r^3}[b(r)-rb'(r)],
\end{equation}
where $F=\frac{df}{dR}$.  The curvature scalar $R$
is given by
\begin{equation}
   R(r)=\frac{2b'(r)}{r^2}.
\end{equation}

For our purposes, a more convenient form of the
line element is
\begin{equation}\label{E:line2}
   ds^2=- e^{\nu(r)} dt^2+e^{\lambda(r)} dr^2
   +r^2( d\theta^2+\text{sin}^2\theta\, d\phi^2).
\end{equation}
Here the Einstein field equations can be written
\cite{fR14}
\begin{equation}\label{E:Einstein1}
8\pi \rho =e^{-\lambda}
\left(\frac{\lambda^\prime}{r} - \frac{1}{r^2}
\right)+\frac{1}{r^2},
\end{equation}

\begin{equation}\label{E:Einstein2}
8\pi p_r=e^{-\lambda}
\left(\frac{1}{r^2}+\frac{\nu^\prime}{r}\right)
-\frac{1}{r^2},
\end{equation}

\noindent and

\begin{equation}\label{E:Einstein3}
8\pi p_t=
\frac{1}{2} e^{-\lambda} \left[\frac{1}{2}(\nu^\prime)^2+
\nu^{\prime\prime} -\frac{1}{2}\lambda^\prime\nu^\prime +
\frac{1}{r}({\nu^\prime- \lambda^\prime})\right] .
\end{equation}
Then if $\nu'\equiv 0$, Lobo's equations become
\begin{equation}\label{E:Einstein1}
   8\pi\rho(r)=
   F(r)\left[e^{-\lambda}\left(\frac{\lambda'}{r}
   -\frac{1}{r^2}\right)+\frac{1}{r^2}\right],
\end{equation}
\begin{equation}\label{E:Einstein2}
   8\pi p_r(r)=F(r)\left[e^{-\lambda}\frac{1}{r^2}
   -\frac{1}{r^2}\right]
   +\frac{F'(r)}{2}\lambda'e^{-\lambda}
      -F''(r)e^{-\lambda},
\end{equation}
and
\begin{equation}\label{Einstein3}
   8\pi p_t(r)= -\frac{F'(r)}{r}e^{-\lambda}
   -\frac{F(r)}{2r}\lambda'e^{-\lambda}.
\end{equation}

Now substituting into the equation of state
$p=\omega\rho$, we obtain
\begin{equation}
   \omega F(r)\left[e^{-\lambda}
   \left(\frac{\lambda'}{r}-\frac{1}{r^2}
   \right)+\frac{1}{r^2}\right]\\
   =F(r)\left[\frac{e^{-\lambda}}{r^2}
   -\frac{1}{r^2}\right]+\frac{F'(r)}{2}\lambda'
   e^{-\lambda}-F''(r)e^{-\lambda}.
\end{equation}
This equation can be rewritten as follows:
\begin{equation}\label{E:F(r)1}
   F''(r)-\frac{1}{2}F'(r)\lambda'+F(r)
   \left[\omega\frac{\lambda'}{r}+(\omega +1)
   \frac{1}{r^2}(e^{\lambda}-1)\right]=0.
\end{equation}

Since we are dealing with a cosmological setting,
we may assume the FLRW model, so that $\nu\equiv 0$:
\begin{equation}\label{E:line1}
ds^{2}=-dt^{2}+a^2(t)\left[\frac{dr^2}{1-kr^2}
+r^{2}(d\theta^{2}+\text{sin}^{2}\theta\,
d\phi^{2})\right].
\end{equation}
Observe that we now have
\begin{equation}\label{E:lambda1}
   e^{\lambda}=a^2(t)\frac{1}{1-kr^2}
\end{equation}
and
\begin{equation}\label{E:lambda2}
   \lambda =\text{ln}\,a^2(t)
        +\text{ln}\,(1-kr^2)^{-1},
\end{equation}
so that
\begin{equation}\label{E:lambda3}
   \lambda'=\frac{2kr}{1-kr^2},
\end{equation}
which is independent of time.  The significance of
the special value $\omega =-1$ in Eq.
(\ref{E:F(r)1}) now becomes apparent: the entire
equation has become time independent, i.e.,
\begin{equation}\label{E:F(r)2}
   F''(r)-\frac{kr}{1-kr^2}F'
       -\frac{2k}{1-kr^2}F=0.
\end{equation}
(For later reference, observe that if $k=0$, then
$F(r)=c_1+c_2r$.)  The solution of Eq.
(\ref{E:F(r)2}) is
\begin{equation}
   F(r)=c_1\,\text{cos}[\sqrt{2}\,\text{ln}\,
     (|k|r+\sqrt{k^2r^2-1})]
   +c_2\,\text{sin}[\sqrt{2}\,\text{ln}\,
     (|k|r+\sqrt{k^2r^2-1})], \quad k\neq 0.
\end{equation}
This solution can also be written
\begin{equation}\label{E:solution1}
   F(r)=c\,\text{sin}[\sqrt{2}\,\text{ln}
   \,(|k|r+\sqrt{k^2r^2-1}+\phi],
\end{equation}
where $c=\sqrt{c_1^2+c_2^2}$ and $\phi=
\text{tan}^{-1}(c_1/c_2)$.
%END OF SECTION

\section{Staying close to Einstein gravity}

In a cosmological setting, $f(R)$ modified gravity
must remain close to Einstein gravity to be
consistent with observation.  In this section
we wish to show that it is possible, at least
in principle, to choose the arbitrary constants
in Eq. (\ref{E:solution1}) so that this goal is
achieved.

The sinusoidal solution (\ref{E:solution1}) has a
large period and a small slope, especially for
large $r$.  To confirm this statement, observe
that the function
\begin{equation}\label{E:g(r)}
   g(r)=\text{sin}[\sqrt{2}\,\text{ln}
   \,(|k|r+\sqrt{k^2r^2-1}]\sim \text{sin}
   (\text{ln}\,r)
\end{equation}
for large $r$.  So both $g'(r)$ and $g''(r)$
approach zero as $r\rightarrow\infty$.  As a
result, $\text{sin}(\text{ln}\,r)$ has the
approximate form $ar+b$ on any interval that
is not excessively large, and since the slope
$a$ is small in absolute value, we have
\begin{equation}
   ar+b\approx b, \quad -1\le b\le 1.
\end{equation}

We can now show that it is possible in principle
to choose the arbitrary constants $c$ and $\phi$
in such a way that $F(r)$ remains close to
unity and both $F'$ and $F''$ close to zero on
one complete period.

Let $\phi =0$, so that
\begin{equation}\label{E:solution2}
   F(r)=c\,\text{sin}[\sqrt{2}\,\text{ln}
   \,(|k|r+\sqrt{k^2r^2-1}],
\end{equation}
First observe that $F(r)=0$ whenever
\[
   \sqrt{2}\,\text{ln}\,(|k|r+\sqrt{k^2r^2-1})
       =n\pi
\]
for all integers $n$.  Solving for $r$, we get
\[
   r=\frac{1}{|k|}\text{cosh}\frac{n\pi}{2}.
\]
Now choose a particular $n$ for which $F(r)\ge 0$ on
the interval $[r_1,r_2]$, where
\begin{equation}
  r_1=\frac{1}{|k|}\text{cosh}\frac{n\pi}{2}
  \quad \text{and} \quad
  r_2=\frac{1}{|k|}\text{cosh}\frac{(n+1)\pi}{2}.
\end{equation}
Next, subdivide the interval $[r_1,r_2]$ into
$i$ subintervals $I_i$ each of which is small
enough so that $b$ remains in a narrow range.
Then on each separate subinterval, construct
a tangent line $a_ir+b_i\approx b_i$ near
the midpoint, thereby ensuring that $b_i\neq 0$.
(See Fig. 1.)  So we may now choose $c_i=1/b_i$
\begin{figure}[tbp]
\begin{center}
\includegraphics[width=0.8\textwidth]{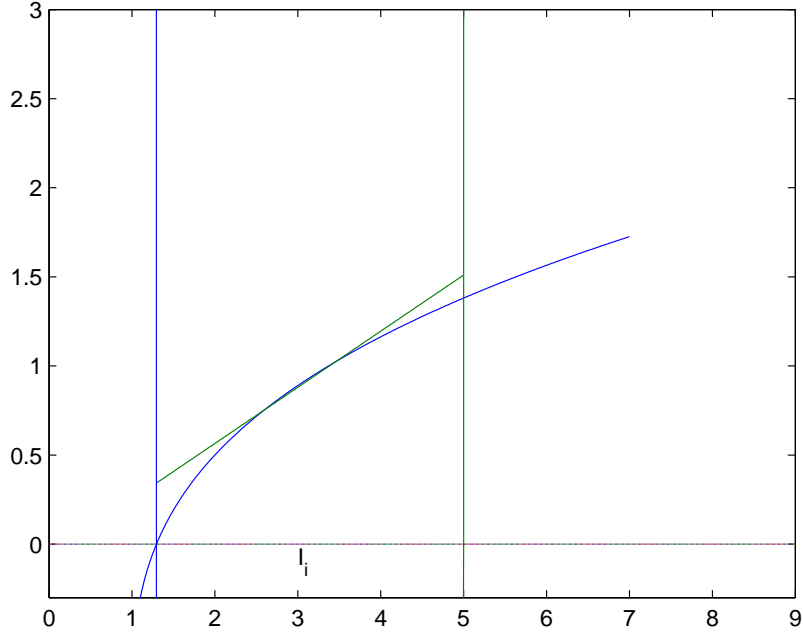}
\end{center}
\caption{The line segment $a_ir+b_i$ on the interval
   $I_i$ (not drawn to scale).}
\end{figure}
for the arbitrary constant $c$.  We then repeat
the procedure on the interval $[r_2,r_2+\pi]$,
so that $F(r)\approx 1$ on the entire period
$[r_1,r_2+\pi]$.  Since both $F'$ and $F''$ are
close to zero [from Eq. (\ref{E:g(r)})], the
periodicity of $F(r)$ guarantees that our
$f(R)$ modified gravity is close to Einstein
gravity for all $r$.
%END OF SECTION

\section{The cosmological constant}
Suppose we return to Eq. (\ref{E:F(r)1}) and
substitute Eqs. (\ref{E:lambda1})-(\ref{E:lambda3}).
Then we obtain
\begin{equation}\label{E:F(r)3}
   F''-\frac{kr}{1-kr^2}F'+F\left[\omega
   \frac{2k}{1-kr^2}+\frac{\omega +1}{r^2}
   \left(a^2(t)\frac{1}{1-kr^2}-1\right)
   \right]=0.
\end{equation}
While we normally assume that $k\neq 0$, it is
noted in Ref. \cite{SF10} that $k=0$,
representing a spatially flat Universe, is
not a dramatic departure from generality
when it comes to late-time cosmology.

With $k=0$, the time-dependent solution is
\begin{multline}\label{E:solution3}
   F(r)=c_1\,\text{exp}[(-\text{ln}\,r)
   (\frac{1}{2}\sqrt{4\omega -4a^2\omega
   -4a^2+5}-\frac{1}{2}]\\
   +c_2\,\text{exp}[(\text{ln}\,r)
   (\frac{1}{2}\sqrt{4\omega -4a^2\omega
   -4a^2+5}+\frac{1}{2})].
\end{multline}
In the special case $\omega =-1$, $F(r)=
c_1+c_2r$, in agreement with Eq.
(\ref{E:F(r)2}) with $k=0$.

In the previous section we dealt with
a time-independent solution due to the
assumption $\omega =-1$.  This allowed
our $f(R)$ modified model to remain
close to Einstein gravity at least in
principle.  By contrast, solution
(\ref{E:solution3}) is time dependent.
So if $\omega \neq -1$, we are dealing
with two possibilities:\\
(a) if $\omega >-1$, there is no real
solution;\\
(b) if $\omega <-1$, then the $f(R)$
model is far removed from Einstein gravity,
i.e., if $a^2(t)$ increases indefinitely,
then the first term in solution
(\ref{E:solution3}) goes to zero, while
the second term gets large.  So $F(r)$
cannot remain close to unity.

We conclude that $\omega =-1$ in the
equation of state $p=\omega\rho$ is the
only allowed value.  Since this note
deals with rather reasonable assumptions,
the only plausible objection to this
conclusion is that the equation of
state for dark energy is much more
complicated than the perfect-fluid
equation of state $p=\omega\rho$.
This possibility was also raised by
Lobo \cite{fL08}, who stated that a
mixture of various interacting
non-ideal fluids may be necessary.
This could imply that dark energy is
dynamic in nature, thereby forcing us
to exclude models with constant $\omega$,
including the cosmological constant.

It should be noted that similar 
conclusions were reached in Ref. 
\cite{CD} using a different and 
somewhat more general approach.
%END OF SECTION

 \section{Conclusion}
 The starting point in this note is
 $f(R)$ modified gravity in a
 cosmological setting.  We also
 assume a spatially flat Universe to
 describe late-time cosmology
 \cite{SF10}; thus $k=0$ in the FLRW
 model.  Our key assumption is the
 perfect-fluid equation of state
 $p=\omega\rho$ to describe the
 hypothesized dark energy.  While
 $\omega <-1/3$ is sufficient to
 yield an accelerated expansion, it
 was concluded that $\omega =-1$ is
 the only value that allows our
 solution to remain close enough
 to Einstein gravity to be
 consistent with observation.

 Weighing the above assumptions,
 we conclude that either (1)
 Einstein's cosmological constant
 is the only acceptable model for
 dark energy or (2) that the
 equation of state is far more
 complicated than the above
 perfect-fluid equation and
 may even exclude a constant
 $\omega$.


\begin{thebibliography}{20}

\bibitem{aR98}Riess, A.G. et al. (1998) Observational
   evidence from supernovae for an accelerating
   universe and a cosmological constant.
   \emph{Astronomical Journal}, \textbf{116}, 1009-1038.
\bibitem{sP99}Perlmutter, S.J. et al. (1999) Measurements
   of $\Omega$ and $\Lambda$ from 42 high-redshift
   supernovae. \emph{The Astrophysical Journal},
   \textbf{517}, 565-586.
\bibitem{SF10}Sotiriou, T.P. and Faraoni, V. (2010)
   $f(R)$ theories of gravity. \emph{Reviews of Modern
   Physics}, \textbf{82}, 451-457.
\bibitem{NO07}Nojiri, S. and Odintsov, S.D. (2007)
   Introduction to modified gravity and
   gravitational alternative for dark energy.
   \emph{International Journal of Geometric Methods
   in Modern Physics}, \textbf{4}, 115.
\bibitem{fL08}Lobo, F.S.N. (2008) The dark side of
   gravity: Modified theories of gravity.
   arXiv: 0807.1640.
\bibitem{rB12}Bousso, R. (2012) The cosmological
   constant problem, dark energy and the landscape
   of string theory. arXiv: 1203.0307.
\bibitem{LO09}Lobo, F.S.N. and Oliveira, M.A. (2009)
   Wormhole geometries in $f(R)$ modified theories
   of gravity. \emph{Physical Review D},
   \textbf{80}, 104012.
\bibitem{fR14}Rahaman, F., Chakraborty, K.,
   Kuhfittig, P.K.F., Shit, G.C., and Rahman, M.
   (2014) A new deterministic model of strange
   stars. \emph{European Physical Journal C},
   \textbf{74}, 3126.
\bibitem{CD}Clifton, T. and Dunsby, P.K.S. (2015) 
   On the emergence of accelerating cosmic 
   expansion in $f(R)$ theories of gravity. 
   \emph{Physical Review D}, \textbf{91}, 
   103528.   

 \end{thebibliography}
\end{document}